\documentclass[journal]{IEEEtran}
\IEEEoverridecommandlockouts
\usepackage{graphicx}
\usepackage{epstopdf}
\usepackage{amsmath}
\usepackage{amssymb}
\usepackage{amsthm}
\usepackage{epsfig}
\usepackage{times}
\usepackage{setspace}
\usepackage{bm}
\usepackage{siunitx}
\usepackage{threeparttable}
\usepackage{balance}
\usepackage{tabularx}
\usepackage{subcaption}
\usepackage{xcolor}
\usepackage{fancyhdr}
\usepackage{diagbox}
\usepackage{array}
\usepackage{float}

\usepackage{textcase}
\usepackage[tablename=TABLE]{caption}
\DeclareCaptionTextFormat{up}{\MakeTextUppercase{#1}}
\usepackage[T1]{fontenc}
\usepackage{times}
\usepackage[mathscr]{eucal}
\usepackage{wrapfig}
\usepackage{multirow}
\usepackage{tablefootnote}
\usepackage{makecell}

\newcommand{\equals}{\!=\!}
\newcommand{\lteq}{\!\le\!}
\newcommand{\minus}{\!-\!}

\newcommand{\gthan}{\!>\!}

\newcommand{\plus}{\!+\!}

\newcommand{\e}{\mathrm{e}}

\newcommand{\dx}{\:\mathrm{d}}

\newcommand{\R}{\mathcal{R}}

\newcommand{\TSP}{\mathcal{S}}

\newcommand{\DC}{\mathrm{DC}}

\begin{document}

% \title{Federated Learning Over LoRaWAN:\\ Simulator Development and Performance Analysis}

\title{Federated Learning Over LoRa Networks: Simulator Design and Performance Evaluation}

\author{Anshika Singh and Siddhartha S. Borkotoky, \IEEEmembership{Member, IEEE} 
\thanks{This work was supported by the Department of Science and Technology (DST), Government of India, through its Core Research Grant (CRG) No. CRG/2023/005803. The authors are with the Indian Institute of Technology Bhubaneswar, Khordha 752050, India. (e-mail: a25ec09003@iitbbs.ac.in, borkotoky@iitbbs.ac.in)}
}

\maketitle

\vspace{-2mm}

\begin{abstract}

Federated learning (FL) over long-range (LoRa) low-power wide area networks faces unique challenges due to limited bandwidth, interference, and strict duty-cycle constraints. We develop a Python-based simulator that integrates and extends the Flower and LoRaSim frameworks to evaluate centralized FL over LoRa networks. The simulator employs a  detailed link-level model for FL update transfer over LoRa channels, capturing LoRa's receiver sensitivity, interference characteristics, block-fading effects, and constraints on the maximum transmission unit. It supports update sparsification, quantization, compression, forward frame-erasure correction (FEC), and duty cycling. Numerical results illustrate the impact of transmission parameters (spreading factor, FEC rate) and interference on FL performance. Demonstrating the critical role of FEC in enabling FL over LoRa networks, we perform an in-depth evaluation of the impact of FEC on FL convergence and device airtime, providing insights for communication protocol design for FL over LoRa networks.
            
\end{abstract}


\section{Introduction}
\label{sec:intro}

Federated Learning (FL) is a distributed machine learning (ML) paradigm in which devices collaboratively train a shared model by iteratively exchanging model parameters rather than raw data. This approach enhances privacy, reduces communication load, and is well suited to Internet of Things (IoT) networks with resource-constrained edge devices.

Long Range Wide Area Networks (LoRaWANs), which use the Long Range (LoRa) physical layer, offer long-distance, low-power communication, making them ideal for battery-powered IoT devices in applications ranging from smart cities to Industry 4.0. Many such applications could benefit from FL, but LoRaWAN’s low data rates (300 bps–27 kbps), strict duty-cycle limits, and interference in unlicensed bands pose challenges that must be evaluated before deployment.

We address this need with a Python-based simulator for FL over LoRa Networks (FLNSim\footnote{https://github.com/ssborkotoky/FLNSim}) that integrates an enhanced LoRa module, built on LoRaSim~\cite{LoRaSim}, into a modified Flower FL framework~\cite{BTM20}. The simulator accepts user-defined network and communication parameters (e.g., network area, interferer density, spreading factor, LoRaWAN class, ping-slot duration, duty cycle) and includes features such as update sparsification, quantization, compression, and forward error correction (FEC). FL update transfers are modeled as LoRa frame transmissions over block-fading channels. To this end, LoRaSim is extended to include inter-spreading-factor interference, fading, duty cycling, and periodic traffic. Outputs include FL performance metrics (accuracy, loss) and communication-cost metrics (uplink/downlink airtime, training time under duty-cycle constraints).

Using MNIST as a case study, we show how spreading factor, FEC rate, and interferer density affect FL convergence and associated communication costs, offering insights for efficient FL deployment over duty-cycled LoRaWANs.

\section{Related Work}

The performance of FL over LoRa networks has been investigated in prior studies~\cite{GFN24,BMN24,SBR25}. In~\cite{GFN24}, a neural-network-based keyword-spotting application is deployed in a federated setting over a LoRa mesh network, with clients executing local ML algorithms on Arduino Portenta H7 microcontrollers. In~\cite{BMN24}, an FL-enabled spectrum-sensing application is designed for LoRa networks and prototyped on the Colosseum network emulator. While both works employ the LoRa physical layer, neither addresses aspects such as duty cycling and operational classes. The impact of duty cycling is considered in the results reported in~\cite{SBR25}, where a federated autoencoder-based anomaly detection framework is proposed. However, the study does not address link-level effects such as fading- and interference-induced frame losses, and thus does not capture their impact on model convergence and communication cost—factors that are critical in unlicensed, interference-prone LoRaWAN environments. Additionally, the studies in~\cite{GFN24,BMN24,SBR25} do not address the issue of frame-erasure correction, which, as we will show later, is imperative for realizing FL over LoRa networks.  

To the best of our knowledge, this work is the first to evaluate FL over LoRa networks while incorporating a detailed link-level model of frame reception that incorporates fading, interference, FEC, and protocol constraints. Moreover, we present a generic and extendable simulation framework that enables systematic investigation of FL performance to help the network designer choose suitable protocol parameters.

\section{Background}
\label{sec:background}

\subsection{Basics of FL}
\label{sec:FL_basics}
Centralized FL involves a server and a number of clients. The clients iteratively train an ML model on their local dataset and send it to the server, which then updates the global model and transmits it back to the clients for further local updates. The key steps in a typical FL routine are as follows:

\begin{enumerate}
    \item  \textit{Initialization:} An initial global model is produced (usually by the server, with random weights and biases)  
    \item \textit{Model Broadcast:} In round $t$, the server broadcasts the current global model parameters $\theta_t$ to the clients.
    
    \item \textit{Local Training:} Client $i$ uses its local dataset $D_i$ to perform model training and computes an updated model  $\theta_t^{(i)}$ (e.g., via stochastic gradient descent).
    
    \item \textit{Update Upload:} Clients send their local updates back to the server.
    
    \item \textit{Aggregation:} The server aggregates the local updates, typically through a weighted-averaging procedure such as federated averaging (FedAvg):
    \begin{align}
        \theta_{t+1} &= \sum_{i=1}^{N} \frac{|D_i|}{\sum_{j=1}^{N} |D_j|} \theta_t^i,
    \end{align}
    where \( |D_i| \) is the size of the local dataset of client \( i \), and $N$ is the number of clients.
    
    \item \textit{Iteration:} Steps 2--5 are repeated over multiple rounds until convergence or a desired performance is achieved.
\end{enumerate}

% \begin{figure*}[t]
%     \centering
% \includegraphics[width=\textwidth]{Figs/FL_diag.pdf}
%     \vspace{2mm}
%     \caption{Key components of the FL process.}
%     \label{fig:FL_diag}
% \end{figure*}

\subsection{Basics of LoRaWAN}
\label{sec:LoRa_basics}

LoRaWAN supports single-hop communication between end devices (EDs) and one or more gateways in both uplink and downlink directions. Data bits are fragmented into frames and transmitted using chirp spread spectrum modulation. A key attribute of a LoRaWAN frame is its spreading factor (SF), an integer between 7 and 12. Larger SFs produce longer frames and therefore require more transmission energy, but they also improve receiver sensitivity, that is, the minimum detectable signal strength decreases as the SF increases. Consequently, larger SFs are typically used for longer links, while smaller SFs are preferred for shorter ones.

The maximum transmission unit (MTU), that is, the maximum payload size in a frame, depends on the SF. For transmissions with a 125-kHz bandwidth over the EU 868 MHz band, the MTU is 222 bytes for SFs 7 and 8, 115 bytes for SF 9, and 51 bytes for SFs 10 through 12. In addition, a maximum duty cycle of 1\% is imposed on transmissions in this band, meaning that over a 100-time-unit interval, a device may transmit for no more than one unit of time.

LoRaWAN EDs typically use pure ALOHA channel access for uplink transmissions. For downlink reception, three operational classes are defined. In Class A, an ED enters receive mode for a short period after each uplink transmission and remains idle otherwise. Class B extends this by adding additional receive windows, called \textit{ping slots}, during which the ED listens for incoming signals. The gateway periodically transmits beacon signals to allow EDs to align their ping slots with the gateway’s schedule. In Class C, the ED remains in receive mode continuously, except during its own transmissions.

\section{Simulation Setup}
\label{sec:sim_setup}
We simulate FL over clients distributed uniformly at random within a circular region, with the server located at the center. The process begins by assigning random weights and biases to the ML model, with all devices initialized to the same model. This can be ensured by setting the random number generator seed identically across all clients. In each round, the server randomly samples a subset of clients for parameter exchange. The server transmits the updated global model to the selected clients, which then compute their local updates and send them back to the server. The server applies federated averaging to the received local updates to derive an updated global model.

The simulation output includes: (i) the accuracy of the global model at the end of each round; (ii) the \textit{round completion time}, defined as the time when all sampled clients finish transmitting their local updates; (iii) the \textit{downlink airtime}, which is the total time the server spends transmitting in a round; and (iv) the \textit{cumulative uplink airtime}, defined as the total time all sampled clients spend transmitting in a round, aggregated across clients.

\subsection{Steps for Update Transfer}
The steps involved in the transmission of an update are described below in the same sequence as they are executed.  

\subsubsection{Sparsification} 
To reduce the communication load, the update may be sparsified by setting all parameter values below a certain magnitude threshold to zero. Currently, fixed thresholding is supported; inclusion of advanced sparsification/pruning techniques is left as future work. 

\subsubsection{Quantization} 
If quantization is enabled, model parameter values are converted to 1, 2, or 4 bit representation. Otherwise, a 32-bit representation is used.

\subsubsection{Compression} 
The global update (possibly sparsified and quantized) is compressed using zlib compression prior to transmission. For further size reduction of local updates, the clients use differential compression strategy in which a client first computes the difference between the local update and the global update. This difference is sparsified, quantized, and compressed using zlib compression.  

\subsubsection{Fragmentation}
To transmit an update of size $U$ bytes (after possible sparsification, quantization, and compression), it is divided into $k$ \textit{source fragments} of $b$ bytes each, where $b$ is the MTU of the SF used for transmission. Thus, 
\begin{align} \label{eq:num_frames}
    k = \left\lceil U/b\right\rceil.
\end{align}

\subsubsection{Forward-Erasure Correction} \label{sec:FEC}
The fragments generated in the previous step can be transmitted as payloads of LoRaWAN frames. However, to mitigate the impact of frame losses, it is desirable to employ FEC. In FEC, the source fragments are encoded using a packet-level erasure-correction code (e.g., MDS code, fountain code, random linear network code) to produce $n \gthan k$ \textit{coded fragments}. The parameter $r \equals k/n$ is referred to as the \textit{rate}. Each coded fragment is a linear combination of the source fragments and has the same length of $b$ bytes. Once the destination receives $k$ linearly independent combinations of source fragments, it can apply a decoding algorithm to recover all $k$ source fragments and reconstruct the update. Our simulator assumes an MDS code, for which any set of $k$ coded fragments suffices for decoding. Thus, an update is considered successfully delivered if at least $k$ coded fragments are received.     

\subsubsection{Frame Transmission}
Each fragment is transmitted as the payload of a LoRaWAN frame. The LoRaSim framework is extended to model frame communications over block Rayleigh-fading channels. For a transmit power of $P_t$ dBm, the received power at distance $d$ is
\begin{align}
    P_r = P_t - L_{\mathrm{ref}} + G_a - \alpha \log_{10}(d/d_{\mathrm{ref}}) + 10\log_{10}A,
\end{align}
where $L_{\mathrm{ref}}$ is the path loss in dB at reference distance $d_{\mathrm{ref}}$~\cite{BRV16}, $\alpha$ is the path-loss exponent, $G_a$ is the system's antenna gain in dB, and $A$ is the power-fading coefficient that is exponentially distributed with mean 1. The transmissions experience interference from interferers distributed according to a Poisson Point Process with intensity $\lambda_I$. Each interferer transmits an average of $\lambda_f$ frames per hour, with the interval between consecutive frames following an exponential distribution. An interfering frame is equally likely to use any of the SFs.

\subsubsection{Downlink Transmission Scheduling}
The transmission of global updates from the server to the clients is performed in the multicast mode. That is, all selected clients attempt to receive the server's transmissions. The transmissions are influenced by the choice of LoRa operation class. The simulator supports
classes B and C. Class A is excluded since it requires each downlink frame to be preceded by an uplink transmission. This is not suitable for supporting the consecutive reception of a sequence of downlink frames.  
For class B LoRaWAN, the starting time of the first frame in an update is aligned with the beginning of a ping slot. In Class C, there is no time slotting because the clients continually listen on the downlink.  

\subsubsection{Uplink Transmission Scheduling}
The local updates are unicast from client to server on the uplink. It is assumed that the number of sampled clients per round is no greater than the number of orthogonal frequency channels available. The clients transmit their updates simultaneously, each using a different frequency band. A processing delay of $\delta_p$ seconds is assumed between the end of the downlink transmission and the start of the uplink phase. Development of advanced scheduling schemes is left for future work.

\subsubsection{Duty-Cycling}
The maximum percentage duty cycle $\DC_{\max}$ determines how frequently a device can transmit updates. First, consider downlink transmissions to \mbox{Class-C} devices. Suppose that the server starts transmitting the update comprising $k$ source fragments of $b$ bytes each using SF $i$ at time $t_0$. Denoting by $l$ the airtime of a frame, the total airtime required to transmit the update is $k l / r$, where $r$ is the FEC rate. To satisfy the duty-cycle constraint, the next update is sent at time $t_0 \plus \Delta$, such that
\begin{align}
\label{eq:interval_C}
    \Delta = \frac{k l}{r \DC_{\max}} \times 100.
\end{align}
For Class B receptions, the interval between the start times of two consecutive updates must be an integer multiple of the ping-slot duration $T_p$ in addition to satisfying~\eqref{eq:interval_C}. Thus, the interval is given by
\begin{align}
\label{eq:interval_B}
    \Delta_B = \lceil \Delta/T_p \rceil \times T_p.
\end{align}
Recall that in round 1, the global model is generated locally at all devices and not actually transmitted. However, an interval of $\Delta$ or $\Delta_B$ is still enforced between rounds 1 and 2 to ensure duty-cycle compliance at the clients.

\subsection{Optional Feature: Semi-Analytical Link Model} \label{sec:analytical_link_model}
To reduce simulation runtime, our program features an option to replace the detailed link-layer simulation with a semi-analytical approximation. Let $\TSP_i(d)$ denote the probability that a frame using SF $i$ is received successfully at a distance $d$. For every frame, a uniform random variable $R$ is produced in the range $[0,1]$. The reception attempt is declared successful if $R \lteq \TSP_i(d)$.  
Using Eq. (14) from~\cite{Bor25}, we can write
\begin{align} \nonumber
\label{eq:avg_TSP}
    &{\TSP}_{i}(d) \approx  \int_{\zeta_i d^{\alpha}/\gamma_0 p_t}^{\infty} \Bigg(1 \minus \\
    &\frac{2 \lambda_f}{\alpha\R_I^2 n_f }\sum_{j=7}^{M}\frac{(l_i(b) \plus \overline{l}_j)\eta_j}{\beta_{j}^{2/\alpha}} \gamma\left(\frac{2}{\alpha},\beta_{j} \R_I^\alpha\right)  \Bigg)^{\overline{n}}\e^{-a}\dx{a},
\end{align}
where $\overline{n} \equals \lambda_I \pi \R_I^2$ is the average number of interferers, $\R_I$ is the distance up to which a device can cause non-negligible interference at the receiver, $\overline{l}_j$ is the average duration of an interfering frame having SF $j$, $\zeta_i$ is the LoRaWAN receiver's sensitivity when using SF $i$, $\xi_{i,j}$ the capture threshold for SF $i$ against interference from SF $j$, $\gamma_0 \equals g_tg_r  L / 4\pi$, \mbox{$\beta_j \equals \xi_{i,j}^{-1}ad^{-\alpha}$}, and $\gamma(i,x) \equals \int_{0}^{x}t^{i-1}\mathrm{e}^{-t}\dx{t}$ is the lower incomplete Gamma function. For more details, the reader is referred to~\cite{Bor25}.

\section{Numerical Results}
\label{sec:results}
We consider the training of an FL model for CNN-based digit classification on the MNIST dataset equally partitioned among 20 clients distributed uniformly over a circle of radius of 500 m.  An 8-channel LoRa link operating in Class B with a ping periodicity of 30 ms is assumed. In each round, 8 clients are randomly sampled for updating. The updates are sparsified with a threshold of 0.001, followed by 4-bit quantization and zlib compression. We assume a transmitter duty cycle of 1\,\%. Unless otherwise specified, the interferers are distributed with intensity \mbox{$10^{-5}$ devices/m\textsuperscript{2}}, each transmitting an average 10 frames per hour. A processing delay of 10 seconds is assumed between the downlink and uplink phases in a round.

\begin{figure}
\begin{subfigure}{0.5\textwidth}
  \centering
  % include first image
  \includegraphics[scale=0.6]{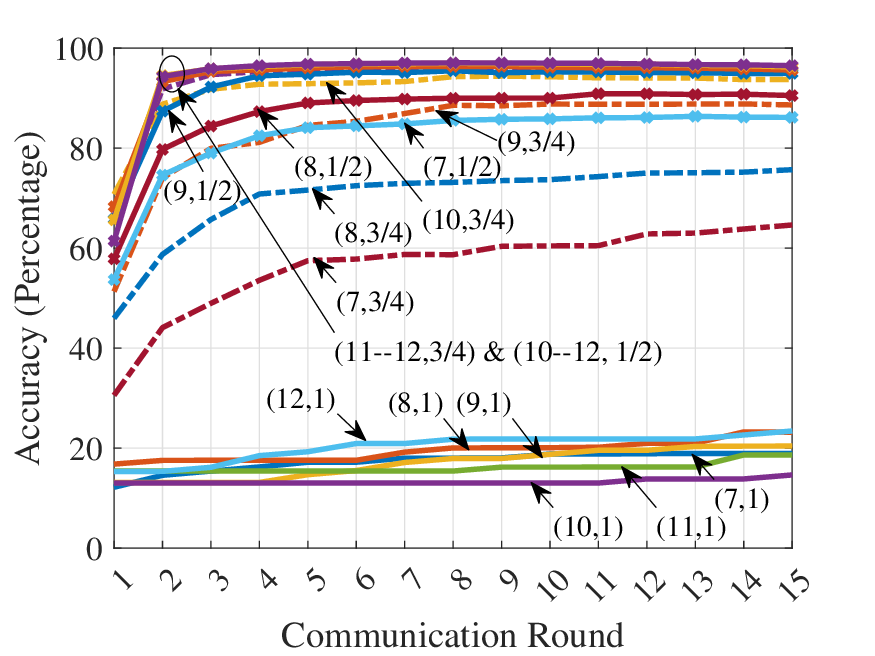}  
  \vspace{-1mm}
  \caption{Accuracy.}
  \label{fig:accuracy}
\end{subfigure}
\begin{subfigure}{0.5\textwidth}
  \centering
  % include second image
  \includegraphics[scale=0.6]{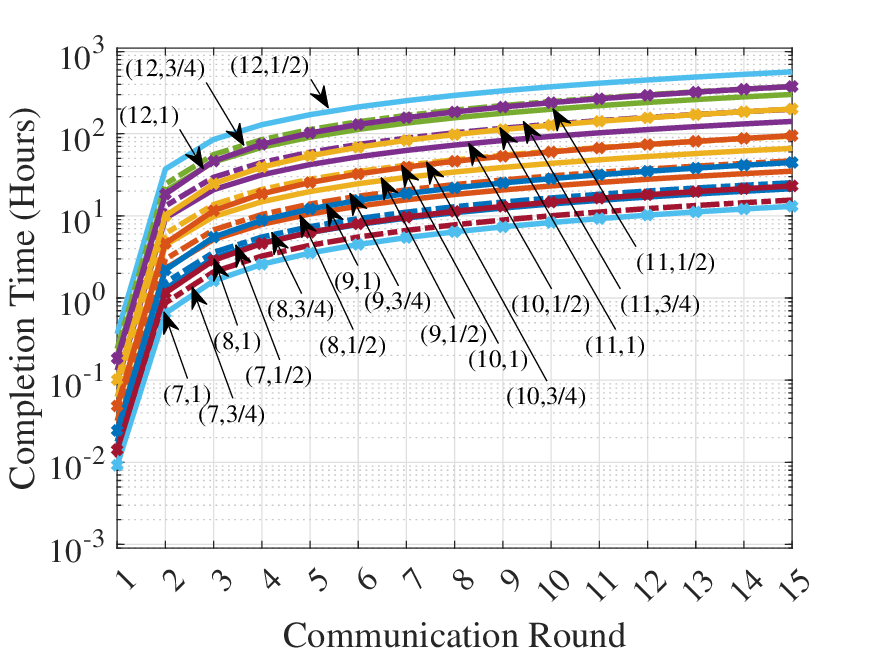}  
  \vspace{-1mm}
  \caption{Completion time.}
  \label{fig:comp_time}
\end{subfigure}
% \vspace{-4mm}
% \setlength\belowcaptionskip{-6mm}
\caption{Accuracy and completion time across rounds.}
\label{fig:acc_and_delay}
\end{figure}

\begin{figure}
\begin{subfigure}{0.5\textwidth}
  \centering
  % include first image
  \includegraphics[scale=0.6]{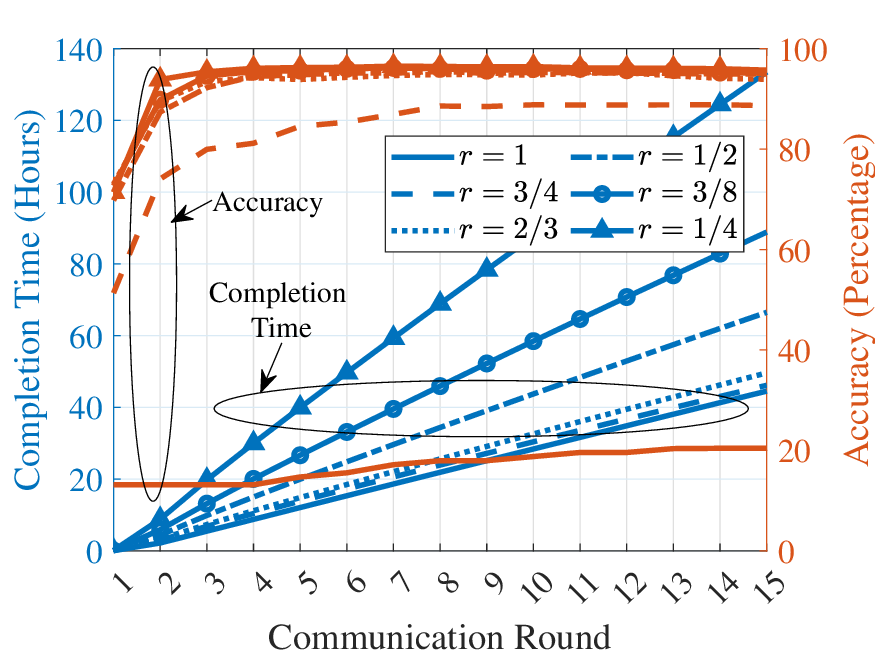}  
  \vspace{-1mm}
  \caption{Accuracy and completion times.}
  \label{fig:FEC_rate_impact}
\end{subfigure}
\begin{subfigure}{0.5\textwidth}
  \centering
  % include second image
  \includegraphics[scale=0.6]{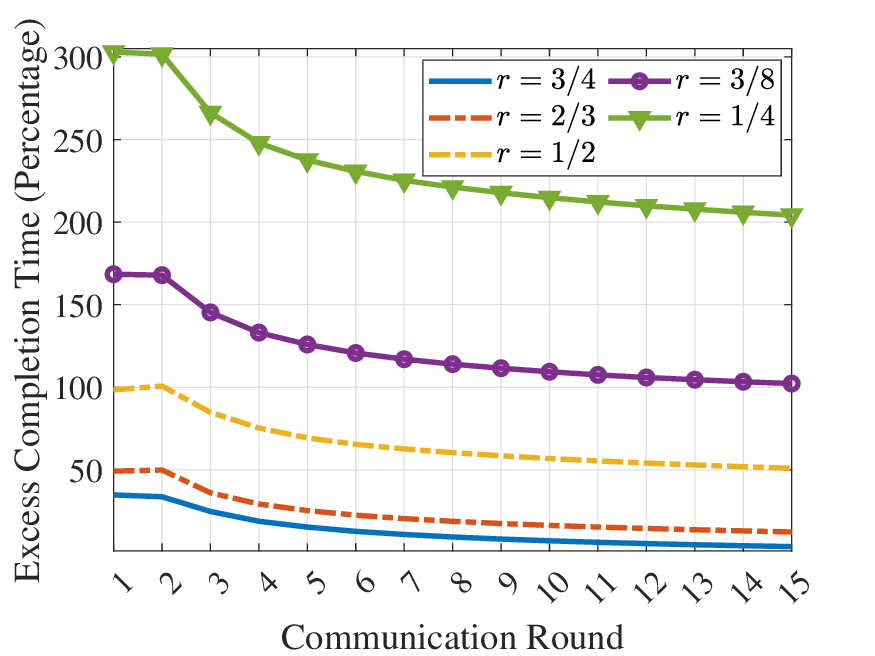}  
  \vspace{-1mm}
  \caption{Percentage increase in the completion time relative to $r\equals 1$.}
  \label{fig:FEC_rate_perc_increase}
\end{subfigure}
% \vspace{-4mm}
% \setlength\belowcaptionskip{-6mm}
\caption{Results for different FEC rates with SF 9.}
\label{fig:acc_and_delay}
\end{figure}

In Fig.~\ref{fig:accuracy}, we plot the percentage accuracy of the global model after each round. The quantities in parentheses specify the SF and the FEC rate $r$. For example, (7,1/2) indicates that the SF is 7 and $r \equals 1/2$. The plot shows that without FEC (i.e., $r \equals 1$), the performance is extremely poor. For any given SF, the use of FEC drastically improves the performance. Furthermore, better performance is achieved by increasing the SF (longer coverage) and decreasing the FEC rate (stronger erasure-correction capability). These gains, however, come at the expense of longer transmission times, since higher SFs yield longer frames and lower FEC rates produce more redundant transmissions. 
These, coupled with duty cycling, result in longer completion times. This is illustrated in Fig.~\ref{fig:comp_time}. 
As expected, the completion time is the smallest for (7,1) and the largest for (12,1/2).

Fig.~\ref{fig:FEC_rate_impact} examines the effect of the FEC rate on the completion time with the SF fixed at 9. A notable observation is the smaller-than-expected relative increase in completion time when the FEC rate decreases. For instance, in round 15, the completion time for $r \equals 1/2$ is about 49\% higher than that for $r\equals 1$. At first glance, this may seem counterintuitive, since transmitting the same update with $r \equals 1/2$ takes roughly twice as long as with $r \equals 1$, suggesting a 100\% increase. Similar discrepancies are seen for other FEC rates.

To investigate this, Fig.~\ref{fig:FEC_rate_perc_increase} plots the percentage increase in completion time relative to a system without FEC. The overhead varies from round to round. In rounds 1 and 2, the excess completion time is approximately $(1-r)/r \times 100 $\%  for each $r$, consistent with the expected increase in transmission time for a given update size when moving from no FEC to rate-$r$ FEC. From round 3 onward, however, the overhead decreases steadily and substantially. This behavior is due to model compression. Initially, the model’s weights and biases are i.i.d. random values, resulting in low compression efficiency. As training progresses and the model approaches convergence, many weights and biases become very small and are subsequently reduced to zero during sparsification. This creates long runs of zeros in the binary representation, which are efficiently compressed by the Lempel–Ziv encoding used in the zlib compression algorithm.

\begin{figure}
\begin{subfigure}{0.5\textwidth}
  \centering
  % include first image
  \includegraphics[scale=0.6]{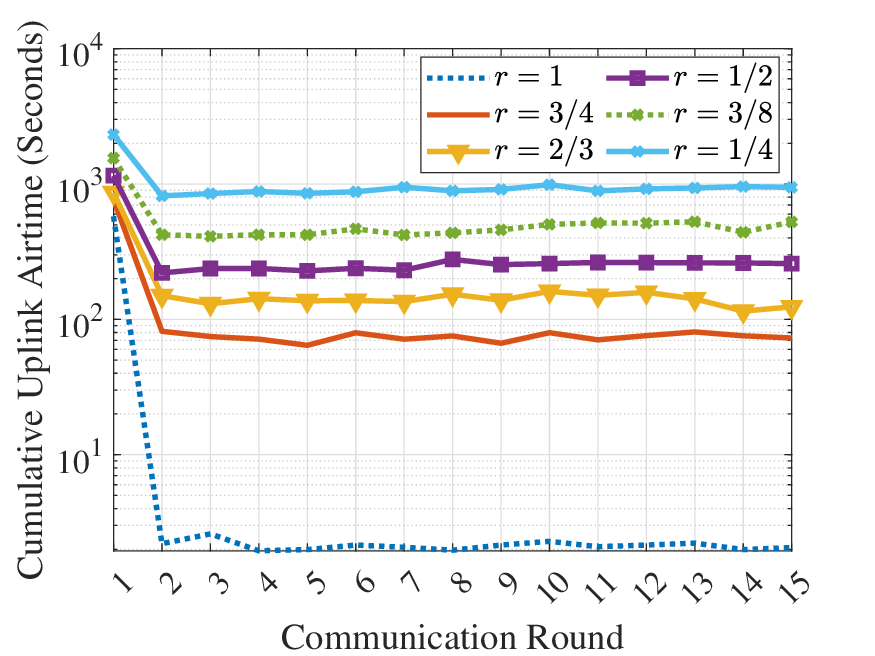}  
  \vspace{-1mm}
  \caption{Uplink airtime.}
  \label{fig:uplink_time}
\end{subfigure}
\begin{subfigure}{0.5\textwidth}
  \centering
  % include second image
  \includegraphics[scale=0.6]{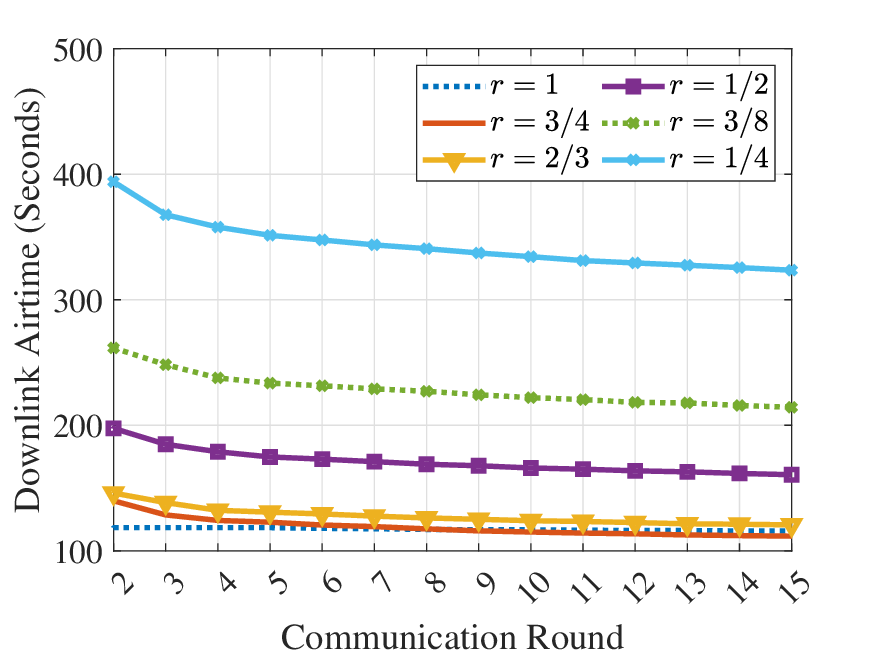}  
  \vspace{-1mm}
  \caption{Downlink airtime.}
  \label{fig:downlink_time}
\end{subfigure}
% \vspace{-4mm}
% \setlength\belowcaptionskip{-6mm}
\caption{Device airtimes for SF 9 and $r \equals 1/2$.}
\label{fig:airtime}
\end{figure}

An analysis of device airtimes sheds light on the energy requirements. Figure \ref{fig:uplink_time} shows the total uplink airtime per round. For any FEC rate, it is highest in round 1, when all clients possess the locally generated global model (Section \ref{sec:sim_setup}) and all sampled clients transmit updates. From round 2 onward, only sampled clients that successfully receive the global model transmit, resulting in reduced airtime. Without FEC ($r \equals 1$), uplink activity beyond round 1 is negligible due to frequent downlink failures. Clients rarely receive the full global update, and hence rarely transmit local updates. 

Figure \ref{fig:downlink_time} shows the downlink airtime. With FEC, it decreases across rounds due to improved compression efficiency as discussed earlier. Without FEC, poor convergence results in an approximately constant communication load. While both uplink and downlink employ compression, the load-variation across rounds is more pronounced on the downlink. On the uplink, the difference between the global and local models often sparsifies into long zero runs, enabling strong compression even in the initial rounds. On the downlink, the raw global model is compressed, with efficiency influenced by its convergence state.

\begin{figure}
    \centering
    \includegraphics[scale=0.6]{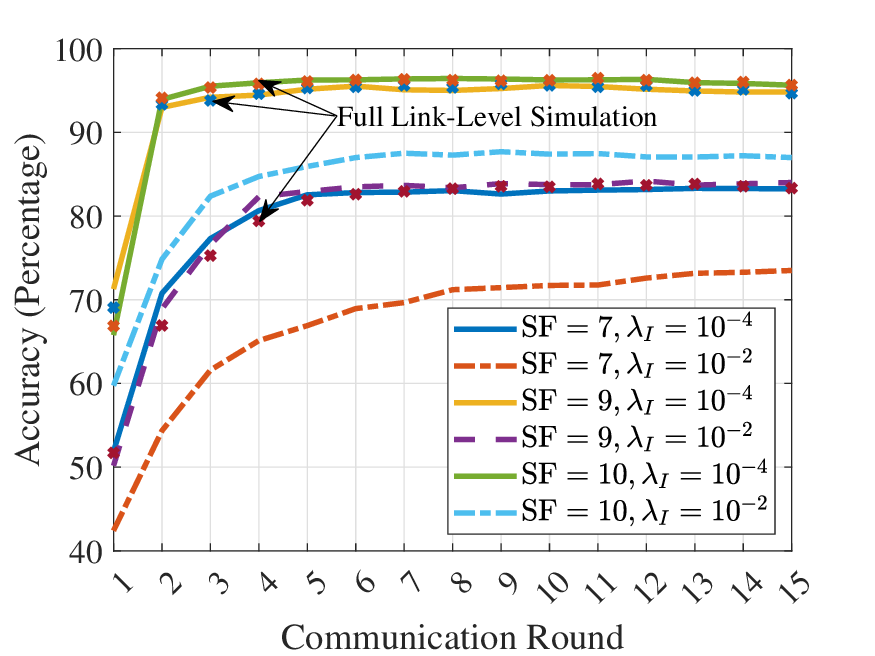}
    \caption{Impact of interference on FL convergence ($r \equals 1/2$).}
    \label{fig:interf_impact}
\end{figure}

Fig.~\ref{fig:interf_impact} shows the impact of interferer intensity $\lambda$ on FL convergence for SFs 7, 9, and 10. A higher interferer intensity results in more collision-induced fragment losses and has a detrimental impact on  FL convergence. Notice that for $\lambda \equals 10^{-4}$, both SFs 9 and 10 provide comparable performance, but for $\lambda \equals 10^{-3}$, SF 10 performs significantly better than SF 9. This is attributed to the stronger interference-rejection capability of the higher SFs. The curves in this plot are obtained via the semi-analytical approach described in Section~\ref{sec:analytical_link_model}. To verify  correctness, we also plot the results obtained by full link-layer simulation (shown as stars) for the case $\lambda \equals 10^{-4}$.

\section{Conclusion}
We developed a simulator for FL over LoRa networks, capturing detailed link-level behavior. Results show that frame losses critically impact FL convergence, making frame-level FEC indispensable. Furthermore, we observe that the per-round communication cost of FL, rather than being a constant, decreases over time as the model approaches convergence. The results also highlight the need for strategies to jointly allocate the SF and FEC rate for optimal performance. The tool provides a practical foundation for designing LoRaWAN-based FL systems and can be extended to incorporate diverse ML models, datasets, and communication-protocol enhancements. 

Future research directions include the development of scalable update scheduling algorithms for FL over a large number of LoRa-enabled clients, development of algorithms for joint allocation of SF and FEC rates for reducing the airtime and latency while maintaining strong convergence performance, and the development of hierarchical aggregation policies for multi-hop LoRa networks for reduced communication range and energy consumption. The simulation platform developed in this work can aid such research.    

%  \vspace*{-2mm}
% \balance
% \bibliographystyle{IEEEtran}
% \bibliography{refs.bib}

\balance
\bibliographystyle{IEEEtran}
\bibliography{refs.bib}

\begin{thebibliography}{1}
\providecommand{\url}[1]{#1}
\csname url@samestyle\endcsname
\providecommand{\newblock}{\relax}
\providecommand{\bibinfo}[2]{#2}
\providecommand{\BIBentrySTDinterwordspacing}{\spaceskip=0pt\relax}
\providecommand{\BIBentryALTinterwordstretchfactor}{4}
\providecommand{\BIBentryALTinterwordspacing}{\spaceskip=\fontdimen2\font plus
\BIBentryALTinterwordstretchfactor\fontdimen3\font minus \fontdimen4\font\relax}
\providecommand{\BIBforeignlanguage}[2]{{%
\expandafter\ifx\csname l@#1\endcsname\relax
\typeout{** WARNING: IEEEtran.bst: No hyphenation pattern has been}%
\typeout{** loaded for the language `#1'. Using the pattern for}%
\typeout{** the default language instead.}%
\else
\language=\csname l@#1\endcsname
\fi
#2}}
\providecommand{\BIBdecl}{\relax}
\BIBdecl

\bibitem{LoRaSim}
{LoRaSim} webpage, \url{https://mcbor.github.io/lorasim/}, Access date: August 7, 2024.

\bibitem{BTM20}
D.~J. Beutel, T.~Topal, A.~Mathur, X.~Qiu, J.~Fernandez-Marques, Y.~Gao, L.~Sani, K.~H. Li, T.~Parcollet, P.~P.~B. De~Gusm{\~a}o \emph{et~al.}, ``Flower: A friendly federated learning research framework,'' \emph{arXiv preprint arXiv:2007.14390}, 2020.

\bibitem{GFN24}
N.~L. Giménez, F.~Freitag, L.~Navarro, and M.~Selimi, ``Demo: Edge federated learning over a {LoRa} mesh network,'' in \emph{2024 IEEE/ACM Symposium on Edge Computing (SEC)}, 2024, pp. 510--511.

\bibitem{BMN24}
F.~Busacca, S.~Mangione, G.~Neglia, I.~Tinnirello, S.~Palazzo, and F.~Restuccia, ``{FedLoRa}: {IoT} spectrum sensing through fast and energy-efficient federated learning in {LoRa} networks,'' in \emph{2024 IEEE 21st International Conference on Mobile Ad-Hoc and Smart Systems (MASS)}.\hskip 1em plus 0.5em minus 0.4em\relax IEEE, 2024, pp. 295--303.

\bibitem{SBR25}
O.~T. Sanchez, G.~Borges, D.~Raposo, A.~Rodrigues, F.~Boavida, and J.~S. Silva, ``Enhancing the performance of industrial {IoT} {LoRaWAN}-enabled federated learning frameworks: A case study,'' \emph{Internet of Things}, p. 101632, 2025.

\bibitem{BRV16}
M.~C. Bor, U.~Roedig, T.~Voigt, and J.~M. Alonso, ``Do {LoRa} low-power wide-area networks scale?'' in \emph{Proceedings of the 19th ACM international conference on modeling, analysis and simulation of wireless and mobile systems}, 2016, pp. 59--67.

\bibitem{Bor25}
S.~S. Borkotoky, ``Balancing the energy consumption and latency of over-the-air firmware updates in {LoRaWAN},'' \emph{IEEE Transactions on Industrial Informatics (Early Access)}, 2025.

\end{thebibliography}

\end{document}